# Key factors of ion induced nanopatterning

S. Bhattacharjee[1], P. Karmakar[2*] and A. Chakrabarti[2]

[1]UGC-DAE Consortium of Scientific Research, III/LB-8, Saltlake, Kolkata -700098, India

[2]Variable Energy Cyclotron Center, I/AF, Bidhannagar, Kolkata -700064, India

We have reported the dependence of projectile mass, chemical reactivity and effect of molecular beams on the ion induced nano structure formation, when 8 keV $He^{1+}$, $N^{1+}$, $O^{1+}$, $Ar^{1+}$ atomic ions and 16 keV $N_2^{1+}$ and $O_2^{1+}$ molecular ions are bombarded on the Si(100) surface at an incidence angle of 60°. Atomic force microscopy (AFM) measurement shows that the initiation and growth of ripple structures are determined not only by the collision cascades but also by the chemical reactivity and molecular state of the projectiles. This experimental investigation explores the necessary requirements for ion induced controlled nanopatterning.

* Corresponding author, E-mail: prasantak@vecc.gov.in

Development of periodic ripple morphology on solid surfaces by oblique incident ion bombardment has become a subject of intense research in recent years because of the controllable wavelength and amplitude of self organized nano patterns [1,2]. This method offers the possibility of producing large-area nano-structured surfaces and has been believed to be an economical and efficient technology for nanostructuring of surfaces.

Formation of ion induced ripple topography depends on the growth or decay of perturbations present on the target surface. The existing models [3-5] based on Sigmund's sputtering theory [6] assumes an ellipsoidal shape of the collision cascade. If the surface has a local curvature, more energy from the collision cascades reaches to the valleys than the hills and therefore the preferential sputtering of the valleys generates instability. The instability combining with thermal diffusion forms the nanostructures. But experimentally, nano structure formation is not observed in all cases except few special cases when beam induced sputtering and diffusion are in favor of such structure formation. Such favorable sputtering and diffusion condition not only depends on the ion energy, fluence, incidence geometry and surface temperature but strongly depends on the chemical and physical character of the target surface. For example, initial roughness present on the surface [7], presence of contamination or multi-element surface leads to easy nanostructure formation [8,9]. Crystallographic orientation and target temperature also control the pattern structures [5]. The variation of projectile on nanostructure formation is thus expected to be important and interesting because projectile mass dependent variation of collision cascade, chemical reactivity of the projectile with target surface and molecular or cluster state of ion may explore interesting fact towards the ion induced nano patterning.

Here we have investigated the role of projectiles of different mass, chemical reactivity and molecular state on nanopatterning of Si (100) by 8 keV $He^{1+}$, $N^{1+}$, $O^{1+}$, $Ar^{1+}$ and 16 keV $N_2^{1+}$ and $O_2^{1+}$ beams. We showed that the variation of projectile mass alter the penetration range and shape of collision cascade that affect the pattern structure; the projectile - surface chemical reactivity generates additional instability (perturbation) on the surface that initiate as well as speed up the nanopattern formation; molecular ion beam leads to early pattern formation than that of atomic ion beam.

Degreased and cleaned Si(100) wafers are bombarded with mass analyzed 8 keV $He^{1+}$, $N^{1+}$, $O^{1+}$, $Ar^{1+}$ atomic ions and 16 keV $N_2^{1+}$ and $O_2^{1+}$ molecular ions at a 60° angle of ion incidence with the surface normal. All the projectiles were bombarded at two fluence, $1\times10^{18}$ atoms/cm$^2$ and $2\times10^{18}$ atoms/cm$^2$. The effective energy for both atomic and molecular ions is same as it is assumed that the molecular beam sputters the surface as two atomic ions of half the incident energy of the molecule. The ion beam was extracted from 6.4 GHz ECR ion source of the Radioactive Ion Beam Facility at Variable Energy Cyclotron Centre Kolkata [10]. For the quantitative morphological analysis the samples were investigated in air by atomic force microscope (AFM) using Nanoscope E.

AFM images of the samples bombarded at an incidence angle of 60° with surface normal by 8 keV $He^{1+}$, $N^{1+}$, $O^{1+}$, and $Ar^{1+}$ ions for the ion fluence of $1\times10^{18}$ ions/cm$^2$ are shown in Fig 1(a), (b), (c), (d) respectively. Similarly, Fig.1(e), (f), (g), and (h) show the AFM images of the same for ion fluence of $2\times10^{18}$ ions/cm$^2$. Figures show that the nanostructures developed on the Si(100) surface are dissimilar for two ion fluences. Moreover, for the same fluence and energy the developed structures differ for different projectiles. AFM topography of the Si surfaces bombarded with $N^{1+}$ and $O^{1+}$ is showing more pronounced ripple structure than the surface bombarded with $Ar^{1+}$ and $He^{1+}$. Ar being inert gas could not develop ripple structures with this combination of fluence and energy whereas nitrogen and oxygen being reactive in nature developed the periodic pattern. Although He is a chemically inert projectile, wavy structure composed of blister like dots are formed.

It is observed that in case of molecular ion bombard the ripples are generated earlier and grow faster compared to that of atomic beam (Fig. 2). Fig. 2(c) and (d) shows the AFM images of the Si surfaces bombarded with 16 keV $N_2^{1+}$ and $O_2^{1+}$ molecular ions, respectively, whereas Fig. 2(a) and (b) showing the same for 8 keV $N^{1+}$ and $O^{1+}$ ion bombardment at same fluence of $2\times10^{18}$ ions/cm$^2$, respectively.

The rms roughness is a measure of height amplitude obtained from the AFM images and the ripple wavelength is defined as the lateral distance between two ripples [11,12]. The variations of the ripple wavelength with mass and molecular state of the projectiles are shown in Fig. 3 (a). The ripple wavelength decreases with the mass of projectile, conversely increases with the longitudinal range of ions. The longitudinal ranges of different projectiles of same energy are calculated using TRIM [13]. It has also been shown earlier that the wavelength increases with the ion range, *a* following an empirical relation $l = 40a$ where $l$ is the ripple wavelength [4,14]. In that case, the range was varied by increasing the ion energy of same projectile.

Fig. 3(b) illustrates the dependence of the rms roughness for different projectiles. With the decrease in projectile mass (increase of longitudinal range), rms roughness increases but for very lighter ions like $He^{1+}$ the trend is not maintained. The ion sputtering is generally determined by atomic processes taking place along a finite range inside the bombarded material. The sputtering increases with the range of the ions but when the range is high enough and the collision cascades go away from the surface, sputtering reduces. Here the rms roughness increases with the increase of projectiles ranges, but as the range is higher in the case of very lighter ions ($He^{1+}$), the collision cascades are formed deep inside the sample leading to less sputtering of surface. Zeberi et al [15] also reported increase of surface roughness for Si and Ge surface with increasing absolute value of ion energy (~ range).

The development of the nanostructure also depends on the reactivity of the projectile ions with the sample. When Si(100) is bombarded with the projectiles like $N^{1+}$ and $O^{1+}$, the initial bombardment develops non uniform chemical phases on the sample surface due to random arrival as well as reactive property of the projectiles ions. Further bombardment leads to the development of nanostructure due to the non uniform sputtering of the same surface because of their compositional change [16]. In case of bombardment of oxygen and nitrogen ions on Si(100) the ripple formation is well-defined compared to Argon and Helium ions. Due to higher reactivity of the $N^{1+}$ and $O^{1+}$ ions, $Si_xN_y$ and $SiO_z$ are formed, respectively, with the initial ion bombardment. The change in chemical composition leads to non uniform sputtering yield of the sample surface which results in perturbation and consequent quick ripple structure formation [16,17]. The FTIRS study of superficial Si layers formed by 9 keV $N_2^+$ ion bombardment demonstrated the existence of $Si_3N_4$ absorption centers [18] and also the formation of $SiO_x$ are shown for 10 keV $O_2^+$ bombardment [19].

$Ar^+$ ions form a damaged layer with the original composition of Si (100). Argon being an inert gas the retention of Ar on Si is much lower as compared to oxygen [16] and nitrogen [17]. K. Elst *et al*. [16] reported that no surface topography is formed under 8 keV $Ar^+$ bombardment while a rough topography is observed under $O_2^+$ bombardment. G. Carter *et al.* [20] observed no waves or ripple trains on Si by 5 or 10 keV $Ar^+$ irradiation at room temperature. In the present case, we observed no well defined ripple topography for 8 keV Ar ion bombardment while reactive projectile forms well defined ripple in this energy regime. The possible reason is that the argon being an inert gas could not form sufficient surface perturbation in this energy regime which is necessary for the ripple formation whereas reactive gases can do by generating different phases on the surface which aids the ripple formation. As well, having the closer mass of oxygen and nitrogen, Neon could not produce any major topographic features even at fluences greater than $10^{20}$ ions/cm$^2$ [21]. However, with further increase of either energy or fluence it is possible to develop ripple structure with the argon ion on Si. We observed ripple formation with 32 keV $Ar^{2+}$ ion bombardment [22]. Chini *et al.* [23] reported ripple formation by 30 and 60 keV Ar ion bombardment on Si at an ion incidence angle of $58^0$ and $60^0$, respectively. Also Carter *et al.* [21] reported ripple like structures on Si at room temperature by 20 keV $Ar^+$ at a very high fluence of $1.7 \times 10^{19}$ ions/cm$^2$.

Although Helium is an inert gas the ripple like structure formation is observed [Fig1.(a),(e)]. With an advantage of blister formation $He^{1+}$ ion could develop nanostructure on Si(100) [24]. Earlier Y. Yamauchi et al. [25] showed the formation of bubble structure on silicon by helium ion bombardment. Thus the formation of bubbles aids to form the observed (Fig.1(a), (e)) ripple like structures by He ion bombardment.

In case of molecular ion bombardment as two atomic ions sputters the surface at same instant the nano structure development is faster (Fig. 2). The observed ripple wavelength and rms roughness (Fig. 3) are also higher in case of molecular oxygen or nitrogen compared to their atomic ions. In case of molecular ions, two atomic projectiles impact the surface simultaneously and the collision cascades generated by the two projectiles overlap and results in nonlinear effects [26]. In case of single atomic ion impact, the chance of simultaneous overlap of two collision cascades for usual ion flux ($<10^{16}$ ions /cm$^2$.sec) is extremely rare [7,27]. The nonlinear effects leads to additional sputtering and surface diffusion which speeds up the ion induced nanostructure formation.

In conclusions, the ripple structure is not formed just by ion bombardment but it depends on the key factors which generate additional instabilities for structure formation on surface. The effect of projectile's mass, chemical reactivity and molecular state are addressed here. Both ripple wavelength and rms roughness decrease with the projectile mass except roughness saturation for He$^{1+}$. The reactive projectiles (N$^{1+}$ and O$^{1+}$) could produce more promising ripple structures at lower fluence compared to the inert projectiles. In the case of molecular ions the ripple structures are formed at lower fluence because of overlapping collision cascades.

The authors would like to thank Dr. V. Ganesan for accessing the SPM, Mr. M. Gangrade for doing SPM measurements and Dr. A. K. Sinha for his constant support.

**Figure Captions**

Figure 1. AFM images of the samples bombarded with 8 keV (a) $He^{1+}$, (b) $N^{1+}$, (c) $O^{1+}$, and (d) $Ar^{1+}$ ions at an angle of $60^0$ for ion fluence of $1\times10^{18}$ ions/cm$^2$. AFM images of the same are shown in (e), (f), (g), and (h) for $He^{1+}$, $N^{1+}$, $O^{1+}$, and $Ar^{1+}$ respectively at an ion fluence of $2\times10^{18}$ ions/cm$^2$.

Figure 2. AFM images of the ripple structures produced by same energy (a) $N^{1+}$ (b) $O^{1+}$ (c) $N_2^{1+}$ (d) $O_2^{1+}$ ions at fluence of $2\times10^{18}$ ions/cm$^2$.

Figure 3. (a) Ripple wavelength (b) Rms roughness vs longitudinal range of projectiles ions. The ion fluences are equal for all cases ($2\times10^{18}$ ions/cm$^2$).

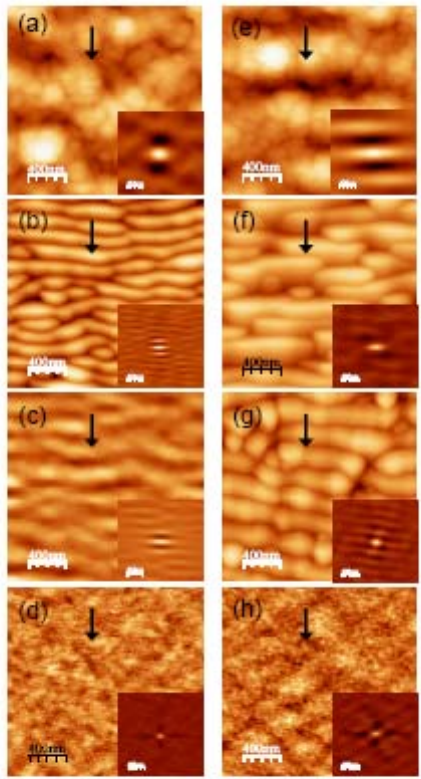

Figure 1

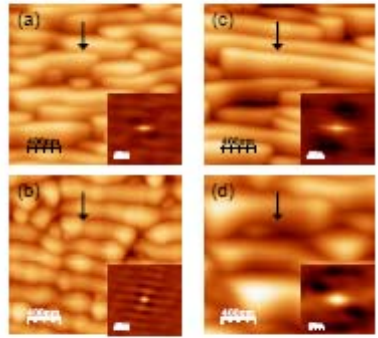

Figure 2

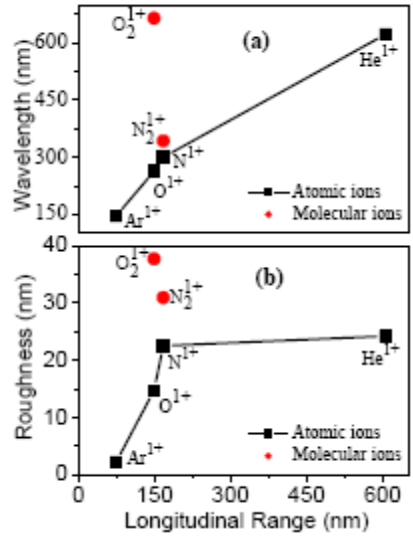

Figure 3